# CMasher: Scientific colormaps for making accessible, informative and 'cmashing' plots

**Ellert van der Velden**[1, 2]

**1** Centre for Astrophysics and Supercomputing, Swinburne University of Technology, PO Box 218, Hawthorn, VIC 3122, Australia **2** ARC Centre of Excellence for All Sky Astrophysics in 3 Dimensions (ASTRO 3D)

# Introduction

The use of colors in the visualization of scientific results is a common sight nowadays. Color allows for more (complex) data to be plotted in the same figure without resorting to difficult-to-interpret 3D plots or subplots; online material; or interactive applications. However, an often underappreciated aspect of data visualization, is how color affects the way the visualized data is interpreted, making it crucial to pick the correct colormap. In order to help with picking a scientific colormap, I introduce the *CMasher* package.

# Background summary

A good scientific colormap is often described/characterized as *perceptually uniform sequential*, which means that the colormap is perceived as uniformly changing in lightness and saturation, mostly at the same hue (Rogowitz, Treinish, & Bryson, 1996; Sharpe, Stockman, Jaegle, & Nathans, 1999). This allows for the data values of a plot to be interpreted correctly by the viewer without giving false information. Such a colormap also often allows for a plot to be converted properly to grey-scale without losing information. A perceptually uniform sequential colormap allows us to properly infer the relative order of the represented numerical values, without requiring a legend or colorbar.

Although there are many works out there that describe the optimal way to do this (Birch, 2012; Brychtová & Çöltekin, 2016; Kindlmann, Reinhard, & Creem, 2002; Rogowitz et al., 1996; Sharpe et al., 1999; Szafir, 2018) and there are tools readily available to test the performance of a colormap (Nuñez, Anderton, & Renslow, 2018; Smith et al., 2018, 2019), bad/misleading colormaps are still very commonly used. The main issue usually is that humans do not perceive every color equally (e.g., small variations in the color green are not perceived as green is a common natural color, while small variations in the colors red and blue are perceived). Here, we use the *jet* colormap to illustrate this issue:

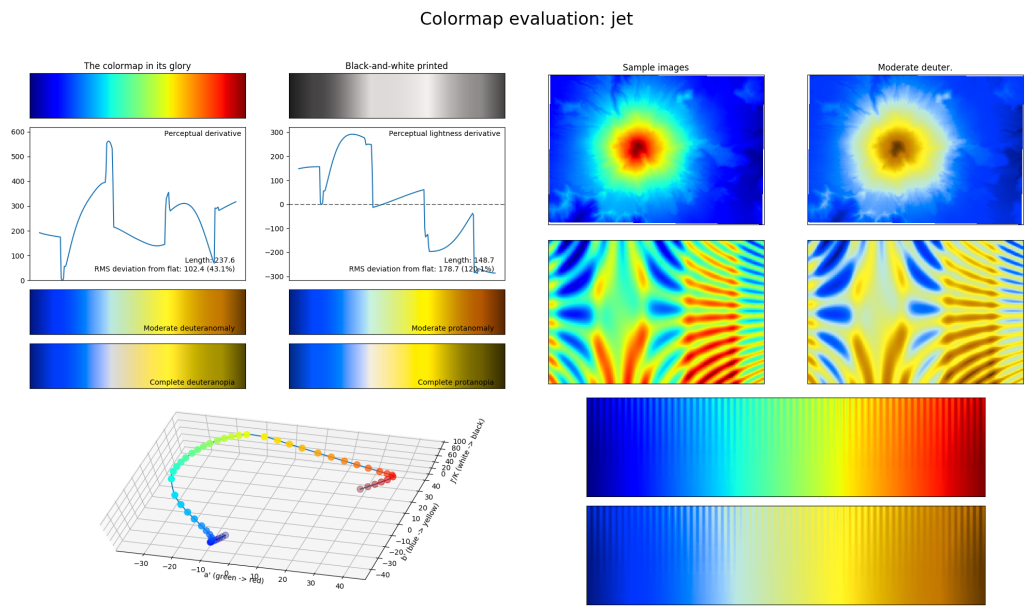

**Figure 1:** Output of the `viscm` package (Smith et al., 2019) showing the statistics and performance of the *jet* colormap. The various different plots show how the colormap changes in perceived saturation and lightness, as well as how well the colormap converts to different types of color-vision deficiency and grey-scale. In case of a perceptually uniform sequential colormap, the two derivative plots should show a straight horizontal line; the colorspace diagram should be smooth; and the lines in the bottom-right corner plots should be visible up to the same depth across the entire colormap.

In Fig. 1, one can view the performance output of the *jet* colormap, made with the `viscm` package (Smith et al., 2019). For perceptually uniform sequential colormaps, the two derivative plots in the top-left should show a straight horizontal line, indicating that the colormap changes uniformly in both perceived saturation and lightness. Consequently, the colorspace diagram in the bottom-left should be smooth. Finally, the lines in the bottom-right plots should be visible up to the same depth across the entire colormap, otherwise it can create artificial features as would be shown by the sample images in the top-right plots. If the colormap is also required to be color-vision deficiency (CVD; color blindness) friendly, the requirements above apply to the deuteranomaly/protanomaly and deuteranopia/protanopia statistics as well.

Using this information, we can check the performance of the *jet* colormap as shown in Fig. 1. The *jet* colormap shows the spectrum of visible light, which trivially increases linearly in wavelength. However, in Fig. 1, we can see that this introduces multiple problems, as the color green is perceived as the brightest of the visible colors due to its natural occurance, and the colormap is absolutely not CVD-friendly. This is an example of a colormap where it would be necessary to have a colorbar/legend, and it is a poor choice for representing numerical values.

Despite all of these shortcomings, *jet* is still a commonly used colormap in the scientific literature. An often cited reason for this (besides the general *"Everyone else uses it."*), is that *jet* has a high perceptual range, making it easier to distinguish adjacent values (*jet* has a higher perceptual range than any colormap in *CMasher*, including the diverging colormaps). Although a high perceptual range can be useful in many different cases, it certainly is not useful in all of them and there are ways to achieve this without giving false information. This is where *CMasher* comes in.

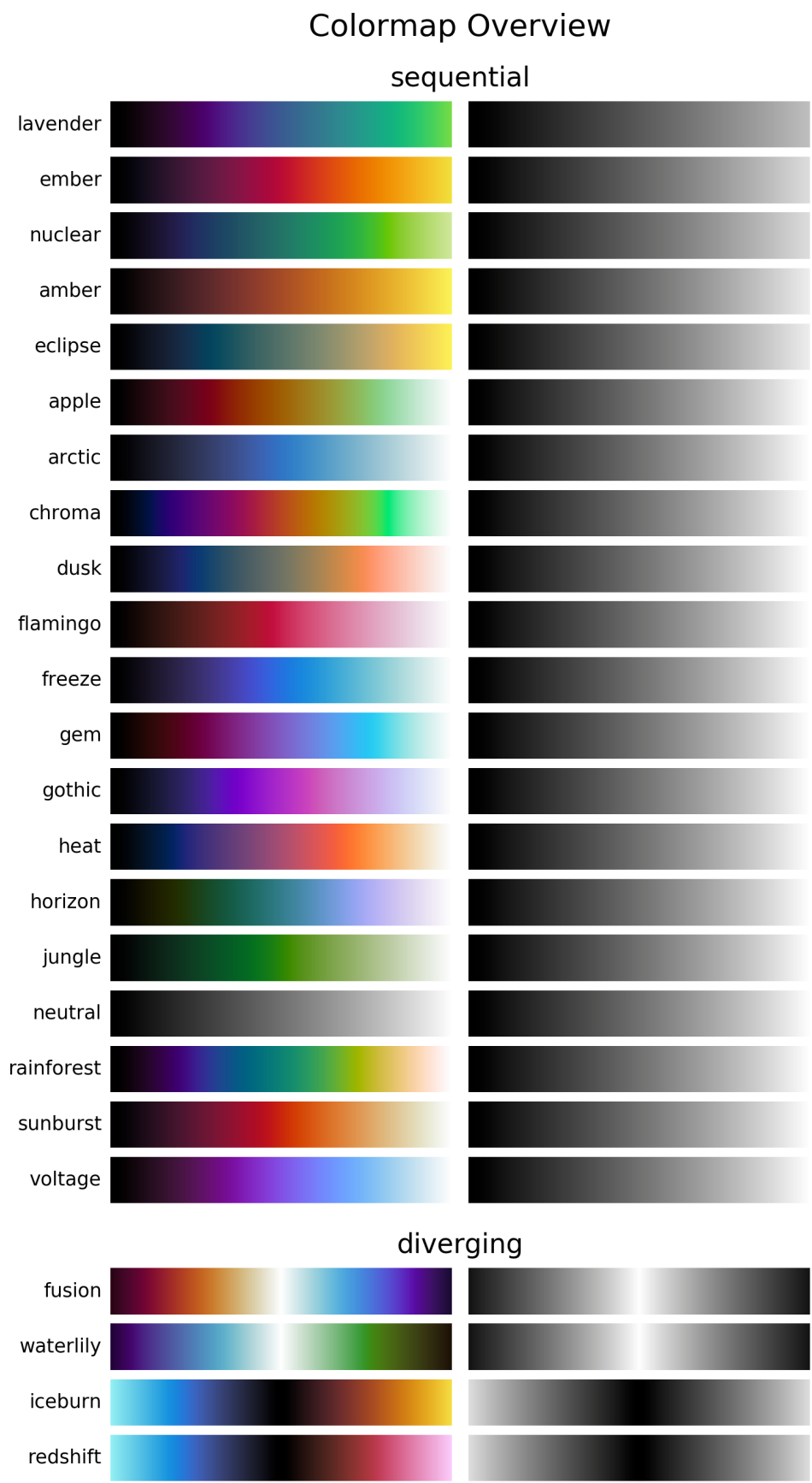

**Figure 2:** Overview of all current colormaps in *CMasher* (v1.2.2).

# CMasher

The *CMasher* package provides a collection of scientific colormaps to be used by different Python packages and projects, mainly in combination with `matplotlib` (Hunter, 2007). The colormaps in *CMasher* are all designed to be perceptually uniform sequential using the `viscm` package (Smith et al., 2019); most of them are CVD-friendly; and they cover a wide range of different color combinations to accommodate for most applications. It offers several alternatives to commonly used colormaps, like *chroma* and *rainforest* for *jet*; *sunburst* for *hot*; *neutral* for *binary*; and *fusion* and *redshift* for *coolwarm*. Users are encouraged to request for specific colormaps to be designed if they cannot find the perfect one. An overview of all current colormaps in *CMasher* (as of v1.2.2) is shown in Fig. 2.

*CMasher* has already been used in several scientific studies, including model emulations (van der Velden, 2019; van der Velden et al., 2019); galaxy kinematics (Džudžar et al., in prep); and redshift estimations for fast radio bursts (Batten, 2019). Due to the number of different color sequences and the perceptual uniform sequential nature of the colormaps, *CMasher* is also great for representing qualitative data. The source code for *CMasher* (including the `viscm` source files) can be found at https://github.com/1313e/CMasher, whereas the descriptions of all available colormaps can be found at https://cmasher.readthedocs.io with their recommended use-cases.

# Acknowledgements

I would like to thank Manodeep Sinha for motivating and inspiring me to make *CMasher*. I would also like to thank Adam Batten; Daniel Berke; Robert Džudžar; and Dexter Hon for their valuable suggestions for improving and expanding *CMasher*. Parts of this research were supported by the Australian Research Council Centre of Excellence for All Sky Astrophysics in 3 Dimensions (ASTRO 3D), through project number CE170100013.